\documentstyle[prd,aps,times,floats,tighten,epsfig]{revtex} 

\newcommand{\Berat}{$^{10}$Be/$^9$Be}
\newcommand{\gray}{$\gamma$-ray\ }
\newcommand{\grays}{$\gamma$-rays\ }
\newcommand{\sigv}{\langle\sigma v\rangle}
\def\Dpp{D_{pp}}
\def\Dxx{D_{xx}}
\def\ddp{{\partial\over\partial p}}

\def\fwa{100 mm}
\def\fwb{180 mm}
\def\fwc{85 mm}
\def\addspace{3\baselineskip}  
\newcommand{\aap}{Astron.\ Astrophys.\ }
\newcommand{\app}{Astropart.\ Phys.\ }
\newcommand{\prep}{Phys.\ Rep.\ }
\newcommand{\plb}{Phys.\ Lett.\ }
\newcommand{\npb}{Nucl.\ Phys.\ }
\newcommand{\pub}[3]{{\bf #1}, #2 (#3)}
\newcommand{\astroph}[1]{{\tt astro-ph/#1}}
\newcommand{\email}[1]{\tt #1}

\begin{document}
\draft

\preprint{1999, PHYSICAL REVIEW D}
\twocolumn 
\wideabs{  

\title{ {\normalsize \rm 1999, PHYSICAL REVIEW D \smallskip }\\
Positrons from particle dark-matter annihilation
in the Galactic halo: \\ propagation Green's functions
}
\author{Igor V.~Moskalenko\cite{imos} and Andrew W.~Strong}
\address{
Max-Planck-Institut f\"ur extraterrestrische Physik,
   Postfach 1603, D-85740 Garching, Germany\\ 
E-mails: \email{imos@mpe.mpg.de, aws@mpe.mpg.de}}

\maketitle

\thispagestyle{empty}  

\begin{abstract}

We have made a calculation of the propagation of positrons from
dark-matter particle annihilation in the Galactic halo in different
models of the dark matter halo distribution using our 3D code, and
present fits to our numerical propagation Green's functions. We show
that the Green's functions are not very sensitive to the dark matter
distribution for the same local dark matter energy density.  We compare
our predictions with computed cosmic ray positron spectra
(``background'') for the ``conventional'' CR nucleon spectrum which
matches the local measurements, and a modified spectrum which respects
the limits imposed by measurements of diffuse Galactic $\gamma$-rays,
antiprotons, and positrons. We conclude that significant detection of a
dark matter signal requires favourable conditions and precise
measurements unless the dark matter is clumpy which would produce a
stronger signal.  Although our conclusion qualitatively agrees with
that of previous authors, it is based on a more realistic model of
particle propagation and thus reduces the scope for future
speculations.  Reliable background evaluation requires new accurate
positron measurements and further developments in modelling production
and propagation of cosmic ray species in the Galaxy.

\end{abstract}
\pacs{95.35.+d, 
      98.35.Gi, 
      98.38.Am, 
      98.70.Sa} 
} 



\enlargethispage*{\addspace}  

\section{Introduction}
\label{sec:intro}

Investigations of galaxy rotation, big-bang nucleosynthesis, and
large-scale structure formation imply that a significant amount of the
mass of the universe consists of non-luminous dark matter
\cite{trimble}.  Among the favored particle dark matter candidates are
so-called weakly interacting massive particles (WIMPs), whose existence
follows from supersymmetric models (see Ref.\ \cite{jkg} for a
review).  If stable, such particles could have a significant
cosmological abundance at the present time.  A pair of stable WIMPs can
annihilate into known particles and antiparticles and it may be
possible to detect WIMPs in the Galactic halo by the products of their
annihilations.  The difficulty, however, consists in discriminating
between the products of WIMP annihilation and ``background'' cosmic ray
(CR) particles.  The smallest background arises when considering
antiprotons and positrons, secondary products of interactions of CR
particles with interstellar matter, and thus these provide the best
opportunity to search for dark matter signatures
\cite{pbarpos,pbar,pos,kamturner}.  A search for a distinct signature
in \grays from the Galactic halo has also been proposed \cite{gamma}.

Though the microphysics is quite well understood and many groups make
sophisticated calculations of the spectra of annihilation products for
numerous WIMP candidates which include many decay chains
\cite{pbar,BE98},  there are still  uncertainties in the macrophysics
which could change the estimated fluxes of WIMP annihilation products
by 1--2 orders of magnitude, making  predictions for their detection
difficult.  In the case of antiprotons the strongest evidence would be
detection of low energy particles \cite{pbarpos,pbar} (below $\sim$1
GeV in the interstellar space) but the solar wind and magnetic field
sweep low energy particles away from the heliosphere, the effect  known
as solar modulation.  In the case of $\gamma$-rays, a weak signal would
compete with the flux of Galactic halo $\gamma$-rays, the uncertain
background of extragalactic photons, and a contribution from unresolved
sources.

The most promising is perhaps the positron signal since it can appear
at high energies where the solar modulation is negligible, but its
strength depends on many details of propagation in the Galaxy.  The
``leaky box'' model is often used \cite{pos,kamturner}, a simplified
approach which may not be applicable in the case of positrons.  The
most accurate propagation model applied so far, the diffusion model
\cite{BE98}, is analytical and thus is subject to certain
simplifications, e.g.\ the positron source function is treated as being
dependent only on the radial cylindrical coordinate $R$, the assumption
of spatially uniform interstellar radiation and magnetic fields, and
some other minor details.  On the other hand, progress in CR positron
measurements is anticipated since several missions operating or under
construction are capable of measuring positron fluxes up to 100 GeV
(e.g.\ gas-RICH/CAPRICE and PAMELA experiments \cite{future_exp}).
Therefore, more accurate calculation of the positron propagation is
desirable.

\enlargethispage*{\addspace}  

We have developed a numerical method and corresponding computer code
(GALPROP)\footnote{Our model including software and datasets is available
at http://www.gamma.mpe--garching.mpg.de/$\sim$aws/aws.html
}
for the calculation of Galactic CR propagation in 3D \cite{SM98}.  The
rationale for our approach was given previously
\cite{SM98,MS98,SMR99,SM99,MSR98,MS99}.  Briefly, the idea is to
develop a model which simultaneously reproduces observational data of
many kinds related to cosmic-ray origin and propagation: directly via
measurements of nuclei, antiprotons, electrons, and positrons,
indirectly via \grays and synchrotron radiation.  These data provide
many independent constraints on any model and our approach is able to
take advantage of this since it aims to be consistent with many types
of observation.

The code is sufficiently general that new physical effects can be
introduced as required.  Its capability includes primary and
secondary nucleons, primary and secondary electrons, secondary
positrons and antiprotons, as well as \grays and synchrotron radiation.
The basic spatial propagation mechanisms are diffusion and convection,
while in momentum space energy loss and diffusive reacceleration are
treated.  Fragmentation, secondary particle production, and energy
losses are computed using realistic distributions for the interstellar
gas and radiation fields.  We aim for a ``standard model'' which can be
improved with new astrophysical input and additional observational
constraints.

In this paper we use our model for calculation of positron propagation
in different models of the dark matter halo distribution. We compare
our predictions with evaluated cosmic ray positron spectra
(``background'') for the ``conventional'' CR nucleon spectrum which
matches the local measurements, and for a modified spectrum which respects
the limits imposed by measurements of diffuse Galactic $\gamma$-rays,
antiprotons, and positrons.  To be specific we will further discuss 
neutralino dark matter, although our results can be easily adopted for
any other particle dark matter candidate.

\section{Basic features of the GALPROP models}
\label{sec:descr}
The GALPROP models have been described in full detail elsewhere
\cite{SM98}; here we just summarize briefly their basic features.

The models are three dimensional with cylindrical symmetry in the
Galaxy, and the basic coordinates are $(R,z,p)$ where $R$ is
Galactocentric radius, $z$ is the distance from the Galactic plane and
$p$ is the total particle momentum. In the models the  propagation
region is bounded by $R=R_h$, $z=\pm z_h$ beyond which free escape is
assumed. 

The propagation equation we use for all CR species is written in the form:
\begin{eqnarray}
\label{eq.1}
\frac{\partial \psi}{\partial t} 
= q(\vec r, p) 
&+& \vec\nabla \cdot ( \Dxx\vec\nabla\psi - \vec V\psi )
+ \ddp\, p^2 \Dpp \ddp\, {1\over p^2}\, \psi \nonumber \\
&-& {\partial\over\partial p} \left[\dot{p} \psi
- {p\over 3} \, (\vec\nabla \cdot \vec V )\psi\right]
- {1\over\tau_f}\psi - {1\over\tau_r}\psi\ ,
\end{eqnarray}
where $\psi=\psi (\vec r,p,t)$ is the density per unit of total
particle momentum, $\psi(p)dp = 4\pi p^2 f(\vec p)$ in terms of
phase-space density $f(\vec p)$, $q(\vec r, p)$ is the source term,
$\Dxx$ is the spatial diffusion coefficient, $\vec V$ is the convection
velocity, reacceleration is described as diffusion in momentum space
and is determined by the coefficient $\Dpp$, $\dot{p}\equiv dp/dt$
is the momentum loss rate, $\tau_f$ is the time scale for
fragmentation, and $\tau_r$ is the time scale for the radioactive
decay. 

An assumption is free escape of particles at the halo boundaries.
Under certain simplifications it translates into the
requirement that the number density of particles at the boundaries is
zero:
\begin{equation}
\label{eq.1.1}
\psi(R_h,z,p) = \psi(R,\pm z_h,p) = 0.
\end{equation}
This is an approximation, but since the number density of particles
there is presumably small, it is resonable to assume that this should
not affect much the particle distribution in the Galaxy.

The numerical solution of the transport equation
(\ref{eq.1})--(\ref{eq.1.1}) is based on a Crank-Nicholson
\cite{Press92} implicit second-order scheme.
Since we have a 3-dimensional $(R,z,p)$ problem we use ``operator
splitting'' to handle the implicit solution.  We apply the
implicit updating scheme alternately for the operator in each dimension
in turn, keeping the other two coordinates fixed.
A check for convergence is performed by computing the timescale
$\psi/ ({\partial \psi/\partial t})$ from eq.~(\ref{eq.1}) and
requiring that this be large compared to all diffusive and energy loss
timescales.
The details of our method are fully explaned in \cite{SM98}.

For a given $z_h$ the diffusion coefficient as a function of momentum
and the reacceleration parameters is determined by CR Boron-to-Carbon
(B/C) ratio.  Reacceleration provides a natural mechanism to reproduce
the B/C ratio without an ad-hoc form for the diffusion coefficient.
The spatial diffusion coefficient is taken as $\beta
D_0(\rho/\rho_0)^\delta$.  Our reacceleration treatment assumes a
Kolmogorov spectrum with $\delta=1/3$.  For the case of reacceleration
the momentum-space diffusion coefficient $D_{pp}$ is related to the
spatial coefficient $\Dxx$ \cite{reacc}.  The injection spectrum of
nucleons is assumed to be a power law in momentum, $dq(p)/dp \propto
p^{-\gamma}$ for the injected particle density, if necessary with a
break.

The total magnetic field is assumed to have the form
\begin{equation}
\label{eq.2}
B_{tot}=B_0\,e^{ - (R-R_\odot)/R_B-|z|/z_B}.
\end{equation}
The values of the parameters ($B_0, R_B, z_B$) are adjusted to match
the 408 MHz synchrotron longitude and latitude distributions.  The
interstellar hydrogen distribution uses HI and CO surveys and
information on the ionized component; the Helium fraction of the gas is
taken as 0.11 by number.  Energy losses for electrons by ionization,
Coulomb interactions, bremsstrahlung, inverse Compton, and synchrotron
are included, and for nucleons by ionization and Coulomb interactions.
The distribution of cosmic-ray sources is chosen to reproduce the
cosmic-ray distribution determined by analysis of EGRET \gray data
\cite{StrongMattox96} and was described
in Ref.\ \cite{SM98}.

Positron production is computed as described in Ref.\ \cite{MS98}; this
includes a critical reevaluation of the secondary decay calculations. 

Gas related \gray intensities are computed from the emissivities as a
function of $(R,z,E_\gamma)$ using the column densities of HI and H$_2$
for Galactocentric annuli based on 21-cm and CO surveys
\cite{StrongMattox96}.  The interstellar radiation field (ISRF), which
is used for calculation of the inverse Compton (IC) emission and
electron energy losses, is calculated based on stellar population
models and COBE results, plus the cosmic microwave background.  Our
results for diffuse continuum $\gamma$-rays, synchrotron radiation, and
a new evaluation of the ISRF are given in
Ref.\ \cite{SMR99}.

An overview of our results is presented in Ref.\ \cite{SM99} and full
results for protons, Helium, positrons, and electrons in
Ref.\ \cite{MS98}.  The evaluation of the B/C and \Berat\ ratios,
evaluation of diffusion/convection and reacceleration models, and full
details of the numerical method are given in Ref.\ \cite{SM98}.
Antiprotons have been evaluated in the context of the ``hard
interstellar nucleon spectrum'' hypothesis in Ref.\ \cite{MSR98}.

\section{Green's functions}
\label{sec:green}
To make a prediction of the positron flux at the solar position, one
needs to know the source function $f(\epsilon)$ which describes the
spectrum of positrons from neutralino annihilation, and the Green's function
$G(E,\epsilon)$ for their propagation in the Galaxy. Then the positron flux
is a convolution
\begin{eqnarray}
\label{eq.3}
\frac{dF}{dE}= \sigv \frac{\rho_0^2}{m_\chi^2}
   \int d\epsilon\, G(E,\epsilon) && \sum_i B_i f_i(\epsilon) \nonumber \\
   && [{\rm cm}^{-2} {\rm\ s}^{-1} {\rm\ sr}^{-1} {\rm\ GeV}^{-1}],
\end{eqnarray}
where $\sigv$ is the thermally averaged annihilation cross section,
$\rho_0$ is the local dark matter mass density, $m_\chi$ is the
neutralino mass, $B_i$ is the branching ratio into a given final state
$i$.  The Green's function thus includes all details of the dark matter
mass distribution and Galactic structure (diffusion coefficient,
spatially and energy dependent energy losses etc.).

\begin{table}
\caption{ Parameters of the dark matter profiles \protect\cite{KK98}.
\label{table1}}
\begin{tabular}{ldd}
Model           & $\rho_0$, GeV cm$^{-3}$ & $r_c$, kpc \\
\tableline
``isothermal''  & 0.43                    & 2.8        \\
Evans           & 0.51                    & 7.0        \\
alternative     & 0.38                    & 0.9        \\
\end{tabular}
\end{table}

\begin{figure}
      \psfig{file=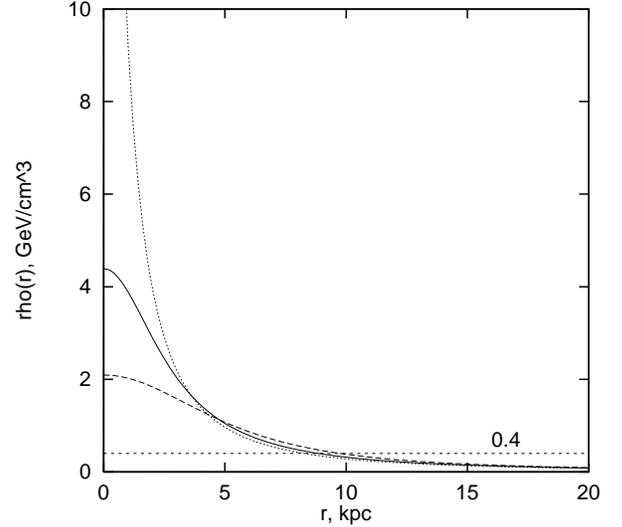,width=\fwa,clip=}
\caption[fig1]{
The radial profiles of the spherical halo models: the canonical
isothermal model (solid line), Evans model (long dashes), alternative
model (dots), and uniform distribution ($\rho = 0.4$ GeV cm$^{-3}$,
short dashes).
}
\label{fig1}
\end{figure}

Following Kamionkowski and Kinkhabwala \cite{KK98} we consider three
different dark matter mass density profiles which match the Galactic
rotation curve. The canonical ``isothermal'' sphere profile,
\begin{equation}
\label{eq.5}
\rho(r) = \rho_0 \frac{r_c^2+R_\odot^2}{r_c^2+r^2},
\end{equation}
where $r_c$ is the core radius, $R_\odot = 8.5$ kpc is the solar
distance from the Galactic center, and $r^2=R^2+z^2$ is the spherical
radial coordinate.  The spherical Evans model,
\begin{equation}
\label{eq.6}
\rho(r) = \rho_0 \frac{(r_c^2+R_\odot^2)^2}{3r_c^2+R_\odot^2}
   \frac{3r_c^2+r^2}{(r_c^2+r^2)^2},
\end{equation}
and an alternative form which also might be empirically acceptable,
\begin{equation}
\label{eq.7}
\rho(r) = \rho_0 \frac{(r_c+R_\odot)^2}{(r_c+r)^2}.
\end{equation}
Note that $\rho_0$ and $r_c$ for each model must be fitted to the
rotation curve, and therefore they are different for each model (see
Table~\ref{table1}).  These profiles are plotted in Fig.\ \ref{fig1}.

For each given model we calculate the Green's function $G(E,\epsilon)$
defined in Eq.~(\ref{eq.3}), which gives the positron flux at the solar
position corresponding to the positron source function in the form of a
Dirac $\delta$-function in energy.  The positron propagation is
calculated in a model which was tuned to match many available
astrophysical data \cite{SM98,SMR99}. Since the halo size in the range
$z_h=4-10$ kpc is favored by our analyses of B/C and \Berat\ ratios and
diffuse Galactic \gray emission \cite{SM98,SMR99}, we consider two
cases $z_h=4$ and $10$ kpc which  provide us with an idea of the
possible limits. The preferred neutralino mass range following from
accelerator and astrophysical constraints is $50$ GeV $< m_\chi <600 $
GeV \cite{ellis98}, and we consider positron energies $\epsilon \le
824$ GeV which cover this range.

For the case of a uniform dark matter mass distribution
$\rho(r)=\left\langle \rho \right\rangle = const$ we compare our
results with simple analytical Green's functions for the leaky-box model
for two cases, where the positron containment time is a constant parameter,
$\tau=\tau_0$,
\begin{equation}
\label{eq.8}
G_1(E,\epsilon) = \frac{c}{4\pi \xi} \frac{1}{E^2} 
   \exp\left(\frac{\epsilon^{-1}-E^{-1}}{\tau_0 \xi}\right)
   \theta(\epsilon-E),
\end{equation}
and when it varies with energy, $\tau(\epsilon)= \eta/\epsilon$,
\begin{equation}
\label{eq.9}
G_2(E,\epsilon) = \frac{c}{4\pi \xi} \frac{1}{\epsilon^2} 
   \left[\frac{E}{\epsilon}\right]^{\frac{1}{\xi\eta}-2}
   \theta(\epsilon-E),
\end{equation}
where $c$ is the  speed of light, $\eta$ is a constant, $\xi$ is the energy
loss constant $d\epsilon/dt = \xi\epsilon^2$, and $\theta(x)$ is the
Heaviside step function.

\begin{figure}
      \psfig{file=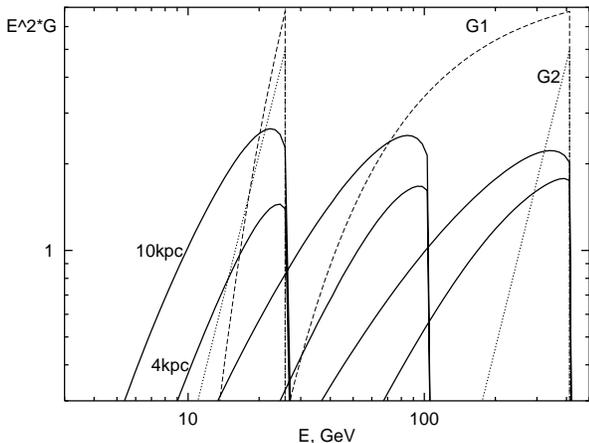,width=\fwc,clip=}
\caption[fig2]{
Calculated $G$-functions for the uniform dark matter distribution,
$z_h=4$ kpc and 10 kpc, for $\epsilon = 25.76$, $103.0$, $412.1$ GeV
(solid lines).  The leaky-box functions $G_1$ and $G_2$ are shown by
dashed and dotted lines respectively.  The units of the abscissa are
$10^{25}$ GeV cm sr$^{-1}$.
}
\label{fig2}
\end{figure}

Fig.\ \ref{fig2} shows functions $E^2 G(E,\epsilon)$ calculated in our
model for  Galactic halo sizes $z_h=4$ and $10$ kpc compared to the
leaky-box functions $G_{1,2}(E,\epsilon)$.  For this comparison we
adopted the following parameters \cite{kamturner}: $G_1$: $\xi =
1.11\times10^{-9}$ yr$^{-1}$ GeV$^{-1}$, $\tau_0=10^7$ yr; $G_2$: $\xi
= 1.52\times10^{-9}$ yr$^{-1}$ GeV$^{-1}$, $\eta=2\times10^8$ yr GeV.
It is clear that the leaky-box model does not work here, moreover a
resonable fit to our $G$-functions is impossible for any combination of
$\xi$ and $\tau_0$ (or $\eta$).  The difference in the normalization at
maximum ($E=\epsilon$) is mainly connected with our accurate
calculation of the ISRF which is responsible for the energy losses.

\begin{figure*}
      \psfig{file=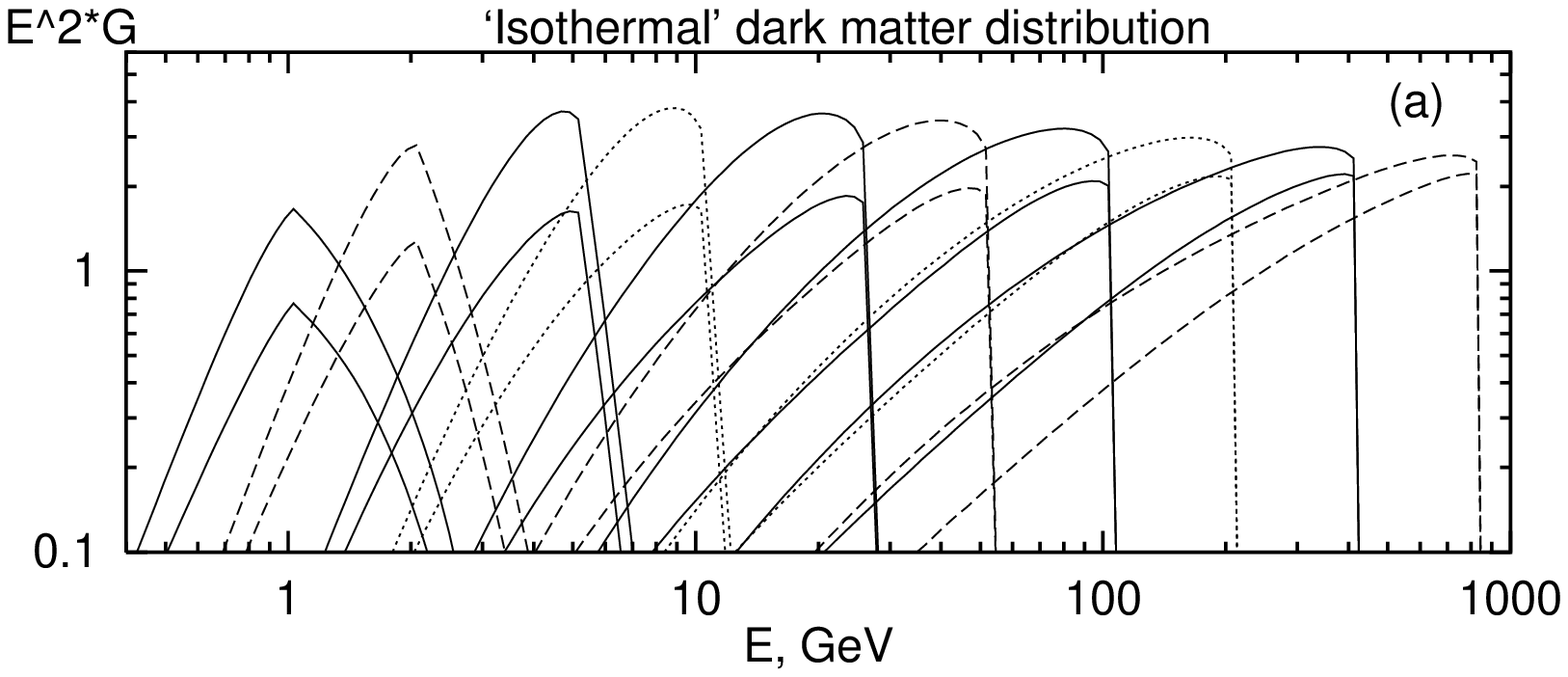,width=\fwb,clip=}\\
      \psfig{file=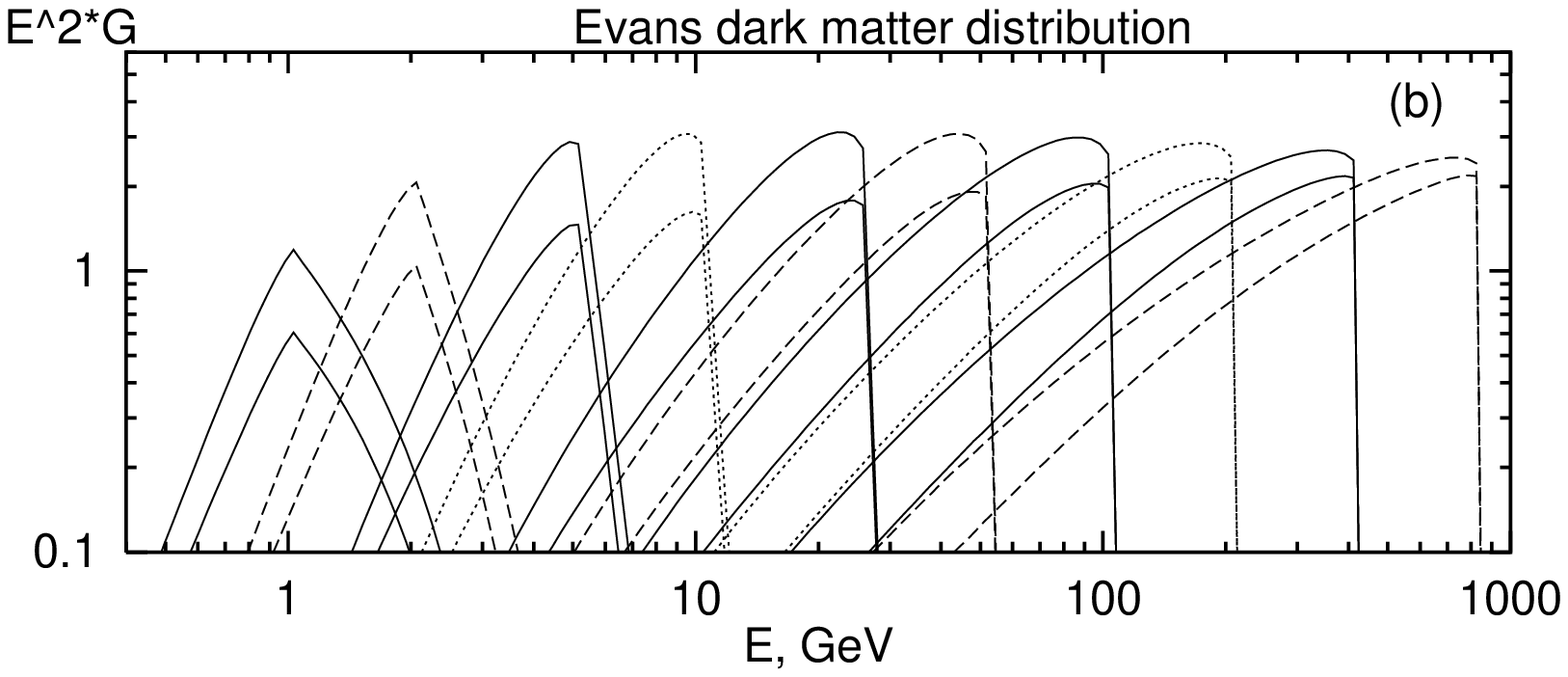,width=\fwb,clip=}\\
      \psfig{file=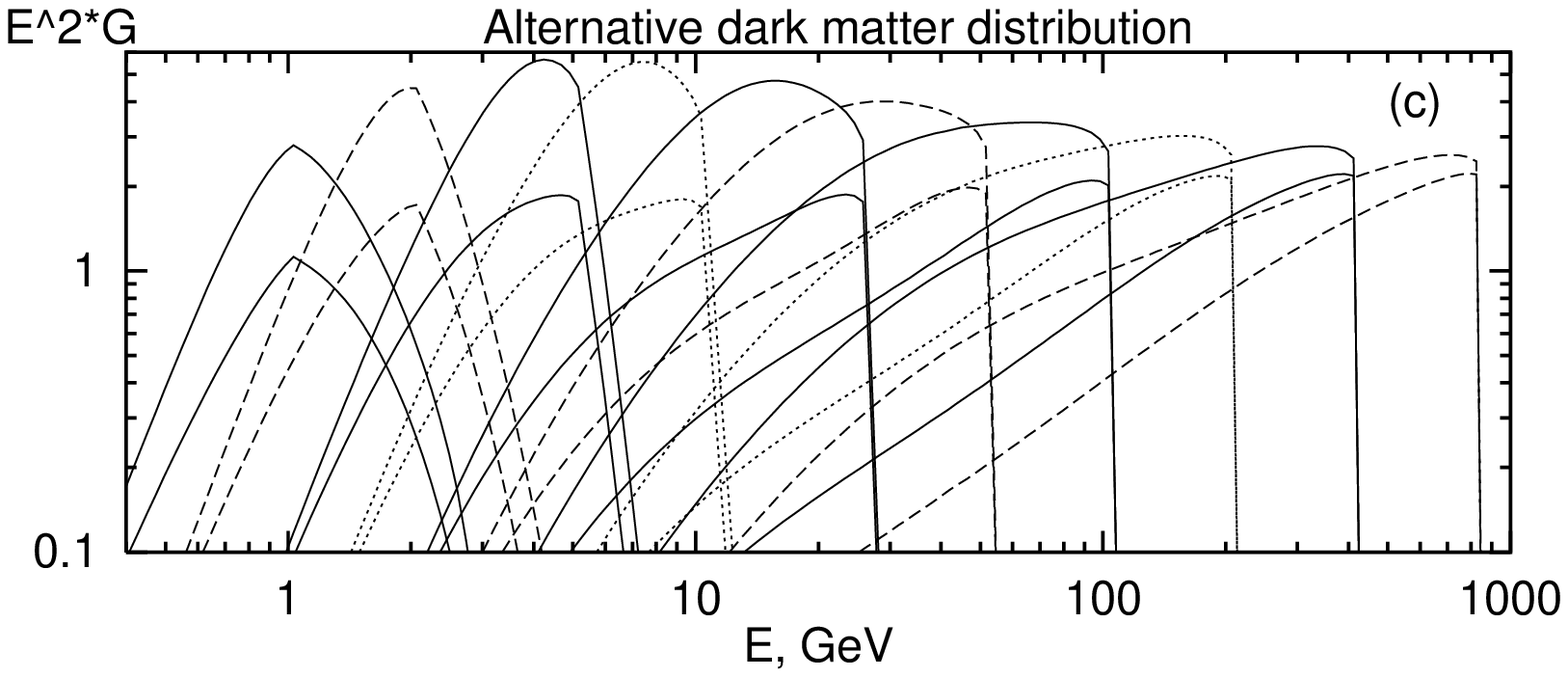,width=\fwb,clip=}
\caption[fig3a,fig3b,fig3c]{
Calculated $G$-functions for different models of the dark matter
distribution: (a) ``isothermal'', (b) Evans, (c) alternative.  Upper
curves $z_h = 10$ kpc, lower curves $z_h = 4$ kpc, $\epsilon = 1.03$,
$2.06$, $5.15$, $10.3$, $25.8$, $51.5$, $103.0$, $206.1$, $412.1$,
$824.3$ GeV.  The units of the abscissa are $10^{25}$ GeV cm sr$^{-1}$.
}
\label{fig3}
\end{figure*}

Fig.\ \ref{fig3} shows our calculated $G$-functions for different models
of the dark matter distribution: ``isothermal'', Evans, and
alternative. The curves are shown for two halo sizes $z_h = 4$ and $10$
kpc and several energies $\epsilon = 1.03$, $2.06$, $5.15$, $10.3$,
$25.8$, $51.5$, $103.0$, $206.1$, $412.1$, $824.3$ GeV.
At high energies, increasing positron energy losses due to the IC
scattering compete with the increasing  diffusion coefficient, while at
low energies increasing energy losses due to the Coulomb scattering and
ionization \cite{SM98} compete with energy gain due to reacceleration.
The first effect leads to a smaller sensivity to the halo size at high
energies.  The second one becomes visible below $\sim 5$ GeV and is
responsible for the appearance of accelerated particles with
$E>\epsilon$.

It is interesting to note that for a given initial positron energy all
three dark matter distributions provide very similar values for the
maximum of the $G$-function (on the $E^2 G(E,\epsilon)$ scale), while
their low-energy tails are different.  This is a natural consequence of
the large positron energy losses.  Positrons contributing to the
maximum of the $G$-function  originate in the solar neighbourhood, where
all models give the same dark matter mass density [see eq.~(\ref{eq.3})
for the definition of the $G$-function].  The central mass density in
these models is very different (Fig.\ \ref{fig1}), and therefore the
shape of the tail is also different since it is produced by positrons
originating in  distant regions.  As compared to the isothermal model,
the Evans model produces sharper tails, while the alternative model
gives more positrons in the low-energy tail.  At intermediate energies
($\sim 10$ GeV) where the energy losses are minimal, the difference
between $z_h=4$ and 10 kpc is maximal.  Also at these energies
positrons from  dark matter particle annihilations in the Galactic
center can contribute to the predicted flux.  This  is clearly seen in
the case of the alternative model with its very large central mass
density (Fig.\ \ref{fig3}c, $z_h=10$ kpc).

To provide the Green's function for an arbitrary positron energy, which
is necessary for prediction of positron fluxes in the case of continuum
positron source functions (as  will be required if one considers secondary,
tertiary etc.\ decay products), we made a fit to our numerical
results.  Since a resonable fit using the leaky-box Green's functions
is impossible we have chosen the function
\begin{eqnarray}
\label{eq.10}
 &&
G(E,\epsilon) = \frac{10^{25}}{E^2}
   \left\{ 10^{a\log^2 E+b\log E+c}
   \,\theta(\epsilon-E)   \right. \nonumber \\
 && \left. 
   +10^{w\log^2 E+x\log E+y}\, 
   \theta(E-\epsilon) \right\}
   \quad[{\rm cm} {\rm\ sr}^{-1} {\rm\ GeV}^{-1}],
\end{eqnarray}
which allows us to fit our numerical functions with accuracy better
than 10\% over a decade in magnitude (on the $E^2 G(E,\epsilon)$
scale).  Here the first term fits the low energy tail, the second
term fits the right-hand-side part of the $G$-functions and represents
the effect of reacceleration, $E$ is in GeV, and
$a(\epsilon),b(\epsilon),c(\epsilon),
w(\epsilon),x(\epsilon),y(\epsilon)$ are the fitting parameters.
Though a better fit is possible by using more complicated functions, we
try to minimize the number of fitting parameters while still providing
resonable accuracy.  Besides, the accuracy of our propagation model is
not better than 10\%, being limited by the accuracy of the
astrophysical data input.  The numerical values of the fitting
parameters are given in Tables~\ref{table2}--\ref{table3} for the
three  models discussed.  At intermediate energies the parameters can
be interpolated.  The cubic spline (or square spline for $w,x,y$)
provides $\sim10$\% accuracy for the $G$-functions when interpolating
on the logarithmic energy scale.

\begin{table*}[t]
\caption{ Fitting parameters of the Green's functions.
\label{table2}}
\begin{tabular}{dddddddddd}
&
\multicolumn{3}{c}{``isothermal'' model} &
\multicolumn{3}{c}{Evans model} &
\multicolumn{3}{c}{alternative model} \\
\cline{2-4} \cline{5-7} \cline{8-10}
$\epsilon$, GeV & $a$ & $b$ & $c$ & $a$ & $b$ & $c$ & $a$ & $b$ & $c$\\
\tableline
\multicolumn{2}{l}{  $z_h = 4$ kpc}\\
  1.03 & $-$1.9732 &    2.3448 & $-$0.1340 & $-$1.9937 &    2.6865 &
 $-$0.2421 & $-$2.0146 &    1.8984 &    0.0364\\
  2.06 & $-$2.4853 &    3.2517 & $-$0.6564 & $-$2.2043 &    3.5919 &
 $-$0.8762 & $-$2.9275 &    2.7980 & $-$0.3409\\
  5.15 & $-$2.6365 &    4.4743 & $-$1.6189 & $-$2.4577 &    4.7878 &
 $-$1.9748 & $-$3.0137 &    4.1174 & $-$1.1397\\
 10.30 & $-$1.9555 &    4.4101 & $-$2.2099 & $-$1.9686 &    4.8817 &
 $-$2.6923 & $-$2.0712 &    3.9274 & $-$1.6167\\
 25.76 & $-$1.1684 &    3.7535 & $-$2.6853 & $-$1.3763 &    4.5614 &
 $-$3.4205 & $-$0.9412 &    2.7981 & $-$1.8075\\
 51.52 & $-$0.8469 &    3.3985 & $-$3.0180 & $-$1.1035 &    4.3731 &
 $-$3.9446 & $-$0.4843 &    2.0882 & $-$1.8434\\
103.00 & $-$0.6979 &    3.3043 & $-$3.4776 & $-$0.9532 &    4.3484 &
 $-$4.5560 & $-$0.2718 &    1.6991 & $-$1.9641\\
206.10 & $-$0.6173 &    3.3187 & $-$4.0118 & $-$0.8078 &    4.1933 &
 $-$5.0264 & $-$0.2586 &    1.8175 & $-$2.4451\\
412.10 & $-$0.5337 &    3.2338 & $-$4.4403 & $-$0.6719 &    3.9382 &
 $-$5.3478 & $-$0.3292 &    2.2664 & $-$3.2998\\
824.30 & $-$0.4255 &    2.9458 & $-$4.6095 & $-$0.5078 &    3.4153 &
 $-$5.2860 & $-$0.3330 &    2.4540 & $-$3.9563\\
\multicolumn{2}{l}{  $z_h =10$ kpc}\\
  1.03 & $-$2.9531 &    2.2688 &    0.2037 & $-$2.2651 &    2.7232 &
    0.0540 & $-$3.9088 &    1.6600 &    0.4355\\
  2.06 & $-$4.3610 &    4.1399 & $-$0.4085 & $-$3.3310 &    4.1276 &
 $-$0.6391 & $-$5.3710 &    4.0111 & $-$0.0642\\
  5.15 & $-$4.6245 &    6.7914 & $-$1.9188 & $-$3.7297 &    6.2467 &
 $-$2.0800 & $-$5.8784 &    7.5056 & $-$1.6622\\
 10.30 & $-$3.6318 &    7.1054 & $-$2.9096 & $-$3.0093 &    6.5384 &
 $-$3.0367 & $-$4.7439 &    8.2050 & $-$2.8244\\
 25.76 & $-$2.3450 &    6.2626 & $-$3.6392 & $-$2.0541 &    5.9838 &
 $-$3.8567 & $-$2.8799 &    6.8568 & $-$3.4165\\
 51.52 & $-$1.6556 &    5.3936 & $-$3.8732 & $-$1.5142 &    5.3295 &
 $-$4.1955 & $-$1.8345 &    5.3634 & $-$3.3196\\
103.00 & $-$1.1707 &    4.5572 & $-$3.9405 & $-$1.1606 &    4.8040 &
 $-$4.4885 & $-$1.1501 &    4.0735 & $-$3.0763\\
206.10 & $-$0.8276 &    3.8034 & $-$3.9013 & $-$0.8889 &    4.2660 &
 $-$4.6500 & $-$0.7058 &    3.0354 & $-$2.7952\\
412.10 & $-$0.5822 &    3.1489 & $-$3.8069 & $-$0.6680 &    3.6993 &
 $-$4.6643 & $-$0.4132 &    2.2082 & $-$2.5103\\
824.30 & $-$0.3984 &    2.5623 & $-$3.6580 & $-$0.4917 &    3.1504 &
 $-$4.5846 & $-$0.2256 &    1.5851 & $-$2.2758\\
\end{tabular}
\end{table*}

\begin{table*}[t]
\caption{ Fitting parameters of the Green's functions.
\label{table3}}
\begin{tabular}{dddddddddd}
&
\multicolumn{3}{c}{``isothermal'' model} &
\multicolumn{3}{c}{Evans model} &
\multicolumn{3}{c}{alternative model} \\
\cline{2-4} \cline{5-7} \cline{8-10}
$\epsilon$, GeV & $w$ & $x$ & $y$ & $w$ & $x$ & $y$ & $w$ & $x$ & $y$\\
\tableline
\multicolumn{2}{l}{  $z_h = 4$ kpc}\\
  1.03 & $-$4.9292 & $-$0.8786 & $-$0.1115 & $-$4.0224 & $-$1.4477 &
 $-$0.2053 & $-$5.9792 & $-$0.1569 &    0.0412\\
  2.06 & $-$7.2475 &    1.0942 &    0.4742 & $-$6.5523 &    0.1919 &
    0.6011 & $-$8.1532 &    2.2559 &    0.3291\\
  5.15 & $-$8.9618 &    2.1785 &    3.1988 & $-$8.0219 &    0.4044 &
    3.9380 & $-$9.3685 &    3.1591 &    2.7479\\
\multicolumn{2}{l}{  $z_h =10$ kpc}\\
  1.03 & $-$4.3201 & $-$1.1227 &    0.2324 & $-$3.5986 & $-$1.4890 &
    0.0867 & $-$5.3468 & $-$0.6192 &    0.4554\\
  2.06 & $-$6.0920 &    0.2541 &    0.9683 & $-$4.2529 & $-$1.4473 &
    1.1879 & $-$8.0942 &    2.1474 &    0.7739\\
  5.15 & $-$6.5457 & $-$1.1929 &    4.7067 &$-$10.0800 &    3.5947 &
    3.0000 & $-$9.0223 &    3.1302 &    3.0000\\
\end{tabular}
\end{table*}

\section{Positron fluxes} \label{sec:pflux}

\subsection{Positrons from the dark-matter particles annihilation}

When neutralinos annihilate in the Galactic halo they produce quarks,
gluons, leptons, and other particles which via hadronization and/or
decays give rise to secondary positrons.  One can expect to get both
monoenergetic positrons (energy $m_{\chi}$) from direct annihilation
into $e^{+}e^{-}$ and continuum positrons from the other annihilation
channels.  In general, the direct $e^{+}e^{-}$ annihilation channel is
severely suppressed with a branching ratio of order $10^{-5}$
\cite{pos}, though some classes of models allow  a larger branching
ratio to be obtained.  Also in some cases, e.g.\ if annihilation occurs
near a pole in the cross section, $\sigv$ can be quite large \cite{res}
which can compensate for a small branching ratio.  The (quasi-)
monoenergetic positron line, if it is strong enough, is easier to
identify.  In contrast, the hadronization and/or decay cascades lead to
the appearance of ``continuum'' positrons with correspondingly degraded
energy thus making worse the signal/background ratio.

For a neutralino heavier then $W^\pm$ ($Z^0$) boson, annihilation to
$W^\pm$- or $Z^0$-boson pairs will be significant followed by the
direct decay of $W^+$'s and $Z^0$'s to lepton pairs where the direct
positron channel accounts for 11\% and 3.4\% of $W^+$ and $Z^0$ decays,
respectively.  In the case where the neutralino is a pure Higgsino state,
the annihilation cross section to $W^\pm$- or $Z^0$-boson pair
increases rapidly above the threshold and reaches a maximum of
$\sigv \approx 3\times 10^{-25}$ cm$^3$
s$^{-1}$ (for $W^+W^-$) and $\approx 2\times 10^{-25}$ cm$^3$ s$^{-1}$
(for $Z^0Z^0$) at about $110$ GeV and $120$ GeV, respectively
\cite{kamturner}.  For unpolarized $W^+$-boson the decay is isotropic
in the rest frame, which produces a uniform positron distribution in
the laboratory system:
\begin{equation}
\label{eq.11}
f(\epsilon)= \frac{1}{m_\chi \beta_W} \,
   \theta(\epsilon-\epsilon_-) \,
   \theta(\epsilon_+ -\epsilon) 
\end{equation}
where $\beta_W$ is the $W^+$-boson speed in the laboratory system,
and $\epsilon_\pm = \frac{1}{2} m_\chi(1\pm\beta_W)$. 
Since the CR positron spectrum falls as $\sim E^{-3.3}$ above
several GeV \cite{Barwick98}, and the signal strength is proportional
to $\epsilon^{-2} m_\chi^{-3}$, the signal/background ratio is maximal
near $m_\chi \sim m_W$.

\subsection{Positron ``background''}

An important issue in interpretation of the positron measurements is
evaluation of the ``background'', positrons arising from CR particle
interactions with interstellar matter.  Though the parameters of the
propagation and the Galactic halo size can be fixed in a
self-consistent way using CR isotope ratios, the ambient CR proton
spectrum on the Galactic scale remains quite uncertain.

The only possibility to trace the spectrum of nucleons on a large scale
is to observe secondary products such as diffuse $\gamma$-rays,
positrons, and antiprotons.  The EGRET data show enhanced
\gray emission above 1 GeV in comparison with calculations based on
locally measured (``conventional'') proton and electron spectra
\cite{hunter97}.  This can be interpreted as implying that the average
spectra of particles in the Galaxy can differ from what we measure
locally, due to details of Galactic structure and, in the case of
electrons, large energy losses.  A possible solution could be a hard
interstellar proton spectrum \cite{HN}, or an electron spectrum which
is on average harder than that locally observed \cite{PP,PE} due to the
spatially inhomogeneous source distribution and energy losses.

The first possibility has been studied in detail in relation to
antiprotons and positrons \cite{SMR99,MSR98}, showing that the
resulting particle fluxes are too large. Taken together, the antiproton
and positron data provide rather substantial evidence against the idea
of explaining the $>$1 GeV \gray excess with a hard nucleon spectrum.

The second possibility looks more plausible and detailed calculations
\cite{SMR99,PE} showed that the \gray excess could
indeed be explained in terms of IC emission from a hard
electron spectrum, but the fit to the EGRET spectral shape is still not
very good.  It can be further improved by allowing some freedom
in the nucleon spectrum at low energies, which is possible since solar
modulation affects direct measurements of nucleons below 20 GeV.
Because of the hard electron spectrum the required modification to the
nucleon spectrum is moderate; as expected the predictions for
antiproton and positron fluxes are larger than in the ``conventional''
model but still within the allowed limits \cite{SMR99}.

In order to show the effect of varying of the ambient proton spectrum,
we compare our results with two models for the CR positron ``background''.
These are a ``conventional'' model (model C) which reproduces the local
directly measured proton and Helium spectra above 10 GeV (where solar
modulation is small), and a model with modified nucleon spectrum (model
HEMN), which is flatter below 20 GeV and steeper above, and arises from
our analysis of Galactic diffuse \gray emission.  The ``background''
spectra are slightly dependent on the halo size.  Since all secondary
particles are produced in the Galactic plane, increasing the halo size
results only in a small decrease of the flux at high energies due to
larger energy losses.  The propagation parameters for these models are
given in Refs.\ \cite{SM98,SMR99}, and the formalism for calculation of
secondary positrons is described in Ref.\ \cite{MS98}.

\subsection{Calculations}

We do not intend to make sophisticated calculations of positron spectra
resulting from numerous decay chains such as best done by, e.g., Baltz
and Edsj\"o \cite{BE98} for many WIMP candidates.  Instead, for
illustration purposes, we simplify our analysis by treating the
annihilation to $W^\pm$ and $Z^0$-pairs.  For $m_\chi < m_W$ we
consider only the direct annihilation to $e^{+}e^{-}$ pairs.  In the
first case we use the cross sections for a pure Higgsino
\cite{kamturner} and the production source function given by
Eq.~(\ref{eq.11}), in the latter case we take $B\cdot \sigv =3\times
10^{-28}$ cm$^3$ s$^{-1}$ and monoenergetic positrons.  These
parameters can be considered as optimistic, but possible
\cite{pos,kamturner}.  To maximize the signal we further choose the
Galactic halo size as $10$ kpc.

\begin{figure}
      \psfig{file=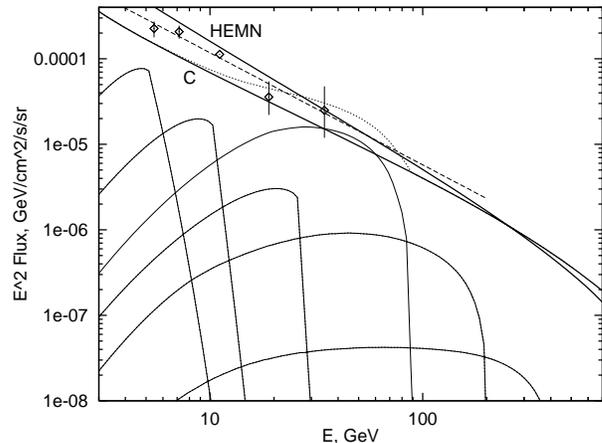,width=\fwc,clip=}
\caption[fig4]{
Our predictions for two CR positron ``background'' models (C and HEMN:
heavy solid lines), and positron signals from neutralino annihilation
for $m_\chi=5.15$, $10.3$, $25.8$, $103.0$, $206.1$, $412.1$ (thin
solid lines):  (a) $z_h=4$ kpc, (b) $z_h=10$ kpc.  In the case of
$m_\chi=103.0$ GeV, the signal plus background (model C) is shown by
the dotted line.  Data and the best fit to the data (dashes) are from
Ref.\ \protect\cite{Barwick98} (HEAT collaboration).
\label{fig4}}
\end{figure}

Fig.\ \ref{fig4} shows our predictions for the two CR positron
``background'' models together with HEAT data \cite{Barwick98} and
positrons from neutralino annihilation. It is seen that the predicted
signal/background ratio has a maximum near $m_\chi \sim m_W$, while
even in the ``conventional'' model the background is nearly equal to
the signal at its maximum.  It is however interesting to note that our
calculations in this model show some excess in low energy ($\leq 10$ GeV)
positrons where the measurements are rather precise but the solar
modulation is also essential. If this excess testifies to a corresponding
excess in  interstellar space and if the positron background
correspond to our ``conventional'' calculations, it could be a hint for
the presence of dark matter \cite{BE98,coutu}.  Our HEMN model fits
the HEAT data better (no excess) and thus provides more background
positrons.
(This shows that in principle a good fit to positron data, which is
consistent also with other measurements such as \grays and antiprotons
is possible without any additional positron source.)
Under such circumstances a significant detection of a weak
signal would require favourable conditions and precise
measurements. Though this our conclusion qualitatively agrees with that
of Baltz and Edsj\"o \cite{BE98} and several earlier papers, it is
based on a more realistic model of particle propagation and thus reduces
the scope for future speculations.

We should mention, however, a possibility which could increase the signal
by orders of magnitude.  Relatively small fluctuations in the dark
matter density distribution will strongly increase the positron yield; this
would be the case if the dark matter halo is clumpy \cite{clumpy}.  But
even if such a signal is detected, its correct interpretation will
require reliable background calculations and thus emphasizes the necessity
for further developments in modelling production and propagation of CR
species in the Galaxy.

\section{Conclusion}
\label{sec:concl}
We have made a calculation of propagation of positrons from  dark-matter
particle annihilation in the Galactic halo using our 3D model which
aims to reproduce simultaneously observational data of many kinds
related to cosmic-ray origin and propagation: directly via measurements
of nuclei, antiprotons, electrons, and positrons, indirectly via \grays
and synchrotron radiation.  We use this model for the calculation of
positron propagation in different models of the dark matter halo
distribution and present fits to our numerical propagation Green's
functions. 

We have shown that the Green's functions are not very sensitive to the
dark matter distribution for the same local dark matter energy density.
This is a natural consequence of the large positron energy losses.  The
differences in the central dark matter mass density lead to  different
shapes of the Green's function low-energy tail, since this involves
positrons originating in  distant regions.  As compared to the
isothermal model, the Evans model produces sharper tails, while
alternative model gives more positrons in the low-energy tail.

We compare our predictions with the computed CR positron ``background''
for two models of the CR nucleon spectrum.  We conclude that a
significant detection of a dark matter signal requires favourable
conditions and precise measurements unless the  dark matter is clumpy
which would produce a stronger signal. Though our result
qualitatively agrees with that of previous authors, it is based on a
more realistic model of particle propagation and thus provides a firmer
basis for conclusions.

A correct interpretation of positron measurements requires reliable
background calculations and thus emphasizes the necessity for further
developments in modelling production and propagation of CR species in
the Galaxy.  The ambient proton spectrum is of primary importance; its
study requires combined approach which allows to evaluate $\gamma$-ray,
antiproton and other data simultaneously.

\acknowledgments
The authors thank the anonymous referee for his valuable comments.

\newpage

\end{document}